\documentstyle[psfig]{EuroPhys}
\input EuroMacr
\begin{document}
\euro{}{}{}{}
\Date{}
\shorttitle{A.P.C.Malbouisson \etal A note on the phase transition...}
\title{\bf A note on the phase transition in a topologically massive 
Ginzburg-Landau theory}

\author{A. P. C. Malbouisson\inst{1}, 
F. S. Nogueira\inst{2}
 and N. F. Svaiter\inst{1}}

\institute{\inst{1} Centro Brasileiro de Pesquisas Fisicas - CBPF,
Rua Dr. Xavier Sigaud 150, Rio de Janeiro, RJ 22290-180, 
BRAZIL\\
\inst{2} Centre de Physique Theorique, Ecole Polytechnique F91128 Palaiseau 
Cedex, FRANCE\\
and\\
Instituto de Matem\'atica e Estat\'{\i}stica - 
Universidade do Estado do Rio de Janeiro - UERJ, Rua S\~ao 
Francisco Xavier 524, Bl. B, Rio de Janeiro, 
RJ 20559-900, BRAZIL}

\rec{}{}

\maketitle

\begin{abstract}
We consider the phase transition in a model which consists of a 
Ginzburg-Landau free energy for superconductors including a Chern-Simons 
term. The mean field theory of Halperin, Lubensky and Ma [Phys. Rev. Lett. 
{\bf 32}, 292 (1974)] is applied for this model. It is found that the 
topological mass, $\theta$, drives the system into different regimes of 
phase transition. For instance, there is a $\theta_{c}$ such that for 
$\theta<\theta_{c}$ a fluctuation induced first order phase transition 
occurs. On the other hand, for $\theta>\theta_{c}$ 
only the second order phase transition exists. The 1-loop renormalization 
group analysis gives further insight to this picture. The fixed point 
structure exhibits tricritical and second order fixed points.
\end{abstract}

The Ginzburg-Landau (GL) theory of superconductivity 
\cite{Ginzburg} describes 
very well the phenomenology of conventional superconductors. 
It is believed that a similar phenomenological theory can be 
applied to the study of the High temperature superconductors 
(HTSC). An important feature which makes the GL theory for HTSC reliable is 
the fact that the critical region in these materials is larger than that of 
the BCS materials.
 However, the GL theory neglects the 
fluctuations of the order parameter which
 are very important in the HTSC. 
Consequently, the exponents of the HTSC phase transition differs 
 from the ones given by the GL theory \cite{Lobb}.
Theoretically, the fluctuations can be taken into account 
through the use of 
renormalization group techniques to study the behavior of the 
theory in the neighbourhood of the critical point 
\cite{HLM,Folk,Berg,Hikami,Kiometzis,Herbut,Lawrie}. 
Another possible path to study the effect of 
the fluctuations is to compute further corrections to the free 
energy functional in a systematic way by performing a loop expansion 
\cite{Kleinert}.  It has been suggested that 
there exists a fluctuation induced first order phase transition in 
superconductors and smectic-A liquid crystals \cite{HLM}. This kind of 
behavior is easily obtained through the fluctuation 
corrected mean field theory of 
Halperin, Lubensky and Ma (HLM) \cite{HLM}. This kind of 
mean field theory relies on 
the observation that the free energy functional is quadratic in the gauge 
fields which allows for a straightforward gaussian integration 
\cite{Malbouisson,Radzi}. For an uniform order parameter, the resulting 
effective free energy exhibits a pattern characteristic of a first order 
phase transition. The renormalization group (RG) analysis gives further 
insight in this picture. The $\epsilon$-expansion performed up to 1-loop 
order in a $N$ component model shows that the superconducting fixed point is 
physically inaccessible for $N<365.9$ \cite{HLM,Lawrie}. This fact has been 
interpreted as a confirmation of the first order phase transition in 
superconductors since in this case $N=2$. Calculations performed up to 
2-loop order does not change this behavior \cite{Folk}. 
The existence of this 
kind of phase transition has been confirmed experimentally in liquid 
crystals where a GL like Hamiltonian is used \cite{exp}. For real 
superconductors, however, we expect a wek first order behavior only in the  
extreme type I regime. For the type II regime a second order 
behavior is expected and the prediction of a first order phase transition in 
this case \cite{HLM} seems to be an artifact of the 
$\epsilon$-expansion \cite{Radzi,Dasgupta}.          

In this note we shall study a generalization of the GL model where a HLM 
like fluctuation corrected 
mean field theory exhibits both first and second order phase 
transition behavior, depending on the physical range of the parameters.  
The model consists of an usual GL free energy functional where a 
topological Chern-Simons (CS) term has been added \cite{Deser}. 
Topological models are frequently employed in the 
construction of quantum models for HTSC which 
explore the effects of statistical transmutation 
(anyon superconductivity) \cite{Fradkin}. Here 
we investigate the effect of such a topological contribution in a 
macroscopic model which generalizes the GL model. 
The resulting fluctuation corrected  
mean field theory is such that the topological mass, $\theta_{0}$, acts as a 
physical control parameter which interpolates between first and second order 
phase transition regimes (in this paper we assume, 
for simplicity, $\theta_{0}\geq 0$, though this is actually not 
necessary). 
This crossover between first and second order 
behavior is characterized by a critical value of the topological mass, 
${\theta_{0}}_{c}$. For $\theta_{0}<{\theta_{0}}_{c}$ we have a first 
phase 
transition while for 
$\theta_{0}>{\theta_{0}}_{c}$ the system undergoes a second order 
phase transition. The perturbative evaluation of the RG functions up to 
1-loop order is consistent with this kind of fluctuation corrected  
mean field behavior. It is obtained that 
the superconducting fixed point is not accessible for a certain range of 
fixed point values of $\mu\theta$ 
(without the subindex $0$, that is, the renormalized parameter. $\mu$ is 
the renormalized magnetic permeability)
while it becomes accessible for another range of $\mu\theta$ fixed 
point values. 
The anomalous dimension for the scalar field 
shows an explicit dependence on the topological mass. A similar behavior is 
obtained in the $CP^{N-1}$ model with a CS term in the context of the $1/N$ 
expansion \cite{Park}. However, this case does not corresponds to a 
genuine GL 
model since a kinetic term for the gauge field is absent. 
See also ref.\cite{Nogueira} for related work in 
a Maxwell-Chern-Simons scalar QED including a $|\psi|^{6}$ term. 

Our starting point 
is the following free energy functional,
 
\begin{equation}
F[\psi,\vec{A}]={\int}d^{3}x\left[|(\nabla-iq_{0}\vec{A})\psi|^{2}
+r_{0}|\psi|^{2}+\frac{u_{0}}{2}|\psi|^{4}
+\frac{1}{8\pi\mu_{0}}(\nabla\times\vec{A})^{2}+i\frac{\theta_{0}}{2}
\vec{A}\cdot(\nabla\times\vec{A})\right],
\end{equation}
where $r_{0}=a_{0}(T-T_{0})/T_{0}$ and $\theta_{0}$ is the topological 
mass. The subindex $0$ in the above parameters denotes bare quantities. 
Since $F$ is quadratic in the vector field, 
the functional integration over 
$\vec{A}$ is Gaussian and can be performed exactly. For 
an uniform $\psi$ this defines the following 
 effective free energy density:

\begin{equation}
f_{eff}[\psi]=-\frac{1}{12\pi}\{[M_{+}^{2}(|\psi|^{2})]^{3/2}+
[M_{-}^{2}(|\psi|^{2})]^{3/2}\}+r|\psi|^{2}
+\frac{u_{0}}{2}|\psi|^{4},
\end{equation}
where the calculations were performed in the the Landau gauge and 
the functions $M_{\pm}$ are defined by

\begin{equation}
M_{\pm}^{2}(|\psi|^{2})=f_{0}|\psi|^{2}+\frac{g_{0}^{2}}{2}\pm
\frac{g_{0}}{2}\sqrt{g_{0}^{2}+4f_{0}^{2}|\psi|^{2}}.
\end{equation} 
where we have defined 
$f_{0}=4\pi\mu_{0}q_{0}^2$ and $g_{0}=4\pi\mu_{0}\theta_{0}$.
The parameter $r_{0}$ has been renormalized to $r=a(T-T_{c})/T_{c}$. 
When $\theta_{0}=0$ Eq.(3) reduces to the effective functional obtained 
by HLM. The critical temperature $T_{c}$ is the same as in the HLM paper. 
 
We have the following phase transition scenario in this model. As 
the temperature is decreased, the system develops a first order 
phase transition with a $f_{0}g_{0}$-dependent 
critical temperature given by

\begin{equation}
T_{c_{1}}=T_{c}\left(1+\frac{f_{0}g_{0}}{2\pi a}\right).
\end{equation}   
In a first order phase transition, there is a local metastable minimum at 
$\psi=0$ and also two local maxima which together with the two global 
minima totalizes five extrema. The local minimum at the origin becomes a 
local maximum when the two local maxima disappears. This will happen at 
the temperature
\begin{equation}
T_{c_{2}}=T_{c}\left(1-\frac{f_{0}g_{0}}{2\pi a}\right).
\end{equation} 
When $g_{0}=0$ 
both critical temperatures coincide and we have  
the fluctuation induced first order transition of HLM.   
An immediate consequence of the above result is the existence of a 
critical value of $g_{0}$ which vanish $T_{c_{2}}$ 
given by ${g_{0}}_{c}=2\pi a/f_{0}$.
This means that for $g_{0}\geq{g_{0}}_{c}$ a second order 
phase transition occurs since there are no local maxima and consequently 
$\psi=0$ will be a minimum only in the disordered phase.
Thus, the introduction of the CS term in the 
GL model implies a very peculiar critical behavior. The topological mass 
drives a crossover between a first and a second order phase transition. 
This is an expected result since for very large values of $\theta_{0}$ the 
gauge modes decouple from the scalar modes and we have in this case the 
limit of a $O(2)$ theory which describes the $\lambda$-transition in liquid 
Helium. Therefore, a crossover region is expected to exist characterized 
by some critical value of $\theta$. We shall see in the next section that 
the effect of the critical fluctuations does not change 
essentially this scenario, though some improvements are necessary. The 
mean field theory discussed in this section is appropriate to describe 
the extreme ``type I'' regime of topologically massive superconductors.  

The critical behavior is better analysed through renormalization 
group (RG) techniques. The case with $\theta_{0}=0$ was already analysed 
by many authors \cite{HLM,Folk,Berg,Hikami,Kiometzis,Herbut,Lawrie}. 
The RG study in the ultraviolet limit 
was carried over in the case of Chern-Simons 
scalar QED without a self-coupling 
of the scalar field and show a trivial behavior of the Chern-Simons 
coupling \cite{Semenoff}.
We are interested in the infrared behavior
 and, therefore, the ultraviolet 
cutoff is kept fixed. We shall work with Wilson's version of the RG in 
its perturbative form \cite{Wilson}. It consists in formally 
integrate out the so called fast modes for the gauge and scalar fields and 
perform the rescalings $\psi\rightarrow e^{t(d+2-\eta)/2}\psi$ and 
$\vec{A}\rightarrow e^{t(d+2-\eta_{A})/2}\vec{A}$ in momentum space. 
Although the presence of the infrared 
cutoff spoils gauge invariance, it can be shown that 
it introduces an anomalous contribution 
to the Ward identities which is well controlled at 
every scale of momenta \cite{Ellwanger}.
 
Let us agree that all paramenters are measured in units of an appropriate 
power of the ultraviolet cutoff $\Lambda$. This can be accomplished by 
putting $\Lambda=1$. The flow equations up to 1-loop order are given by    

\begin{eqnarray}
\frac{dr}{dt}&=&(2-\eta)r+\frac{u}{\pi^{2}}+\frac{f}{\pi^{2}(1+g^{2})},
 \\
\frac{du}{dt}&=&(1-2\eta)u-\frac{5u^{2}}{2\pi^{2}}-\frac{2f^{2}}{\pi^{2}
(1+g^{2})^{2}},\\
\frac{d\mu}{dt}&=&\mu\left(\eta_{A}-\frac{f}{2\pi^{2}}\right),\\
\frac{dq^{2}}{dt}&=&(1-\eta_{A})q^{2},\\
\frac{d\theta}{dt}&=&(1-\eta_{A})\theta,\\
\frac{df}{dt}&=&f-\frac{f^{2}}{2\pi^{2}},\\
\frac{dg}{dt}&=&g\left(1-\frac{f}{2\pi^{2}}\right),
\end{eqnarray}
where 

\begin{equation}
\eta=-\frac{2f}{3\pi^{2}(1+g^{2})}.
\end{equation}
We observe that the fixed point structure is completely independent of 
$\eta_{A}$ just like in the case of the ordinary Ginzburg-Landau model 
\cite{HLM}. Therefore we can fix $\eta_{A}$ at its fixed point value, that is, 
$\eta_{A}=1$ in such a way that there is no flow to $q^{2}$ and $\theta$.
Note that the above flow equations 
were obtained in an uncontrolled approximation since there is no small 
parameter like $\epsilon$ or $1/N$. 

The fixed point structure is well known for $g=0$. 
It is found that the superconducting fixed point has a 
complex value for $u^{*}$ if the number of components of 
the order parameter is less than 365.9 \cite{HLM} which 
is usually interpreted as a signature of the first order phase transition
driven by fluctuations. 
This behavior is changed for  
values of $g>0$. In this 
situation, real superconducting fixed points are found which exhibit 
a stable infrared behavior in the plane $f-u$. The charged fixed 
points are given by $f^{*}=2\pi^{2}$, $g^{*}>0$ arbitrary and 

\begin{equation}
u^{*}_{\pm}=\frac{\pi^{2}}{5(1+{g^{*}}^{2})}\left({g^{*}}^{2}+\frac{11}{3}
\pm\frac{1}{3}\sqrt{\Delta}\right),
\end{equation}
where

\begin{equation}
\Delta=({g^{*}}^{2}-a_{+})({g^{*}}^{2}-a_{-}),
\end{equation}
with $a_{\pm}=-11/3\pm 4\sqrt{5}$. If $g^{*}<\sqrt{a_{+}}$ 
then $\Delta<0$ and 
there is no superconducting fixed point. This situation corresponds to the 
case of a first order phase transition. On the other hand, if 
if $g^{*}>\sqrt{a_{+}}$, we have an infrared stable fixed point corresponding 
to a second order superconducting phase transition. In this situation we 
have also a tricritical fixed point which is over a line of infrared 
attraction (the tricritical line)
which separates the regions of first and second order behaviors. In this case 
the flow diagram is similar to the one found by Herburt and Tesanovic 
\cite{Herbut} and Bergerhoff {\it et al.} \cite{Berg}. Fig. 1 shows the 
flow diagram for the fixed point value $g^{*}=2.5$.  
Note that the 
second order fixed point and the tricritical fixed point will colapse for 
$g^{*}_{c}=\sqrt{a_{+}}\approx 2.2973$ since it implies $\Delta=0$. 
This result is consistent with the 
fluctuation corrected mean field analysis.  

\begin{figure}
\centerline{\psfig{figure=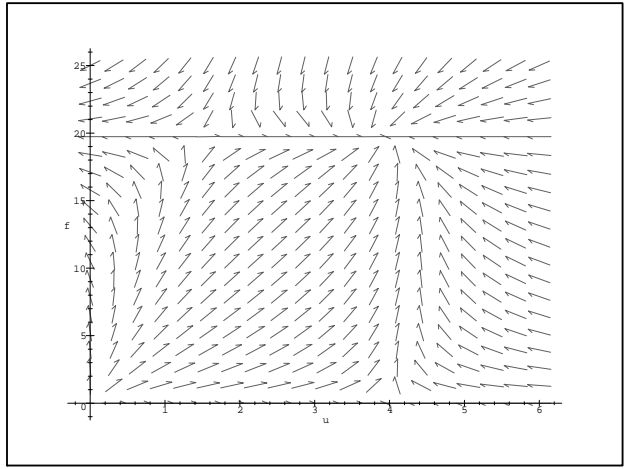,height=7truecm,angle=-90}}
\caption{The tricritical and the second order fixed points for 
$g^{*}=2.5$.}
\end{figure}

Summarizing, 
we have shown in this work how the introduction of a Chern-Simons term in 
the Ginzburg-Landau free energy changes the mean field behavior with 
respect to the conventional situation. The crossover between 
different critical behaviors is manifested due to an explicit dependence 
of the critical temperatures on the CS mass. Also we have computed the 
RG functions up to 1-loop order and verified that there exists an infrared 
stable fixed point for a certain range of values of the CS mass. 
We can ask about the experimental relevancy of the proposed model. 
Probably this model is of 
little interest in what concerns conventional  
superconducting materials. 
However, it can be relevant in the 
study of liquid crystals or even the HTSC. In fact, the study of 
chirality in liquid cristals provides an example where a CS like term is 
added to the free energy \cite{Lubensky}. Also, it is possible to perform 
an appropriate transformation in the Franck free energy to generate a CS 
term \cite{Nogueira2}.
We hope that this work can stimulate new experimental research in the field 
and theoretical investigations as well.  

\stars

F.S.N. would like to thank Prof. C. de Calan for numerous discussions and 
Profs. J. Magnen and 
V. Rivasseau for the kind hospitality at the Centre de Physique 
Theorique - Ecole Polytechnique. A.P.C.M. would like to thank the 
hospitality of CERN - Theory Division, where part of this work has been 
done. This work was supported by the agency CNPq, a division of the 
brazilian Ministry of Science and Technology.

\end{document}